# Efficient, sub-4-cycle, 1-µm-pumped optical parametric amplifier at 10 µm based on BaGa$_4$S$_7$

Zsuzsanna Heiner,[1,*] Valentin Petrov,[2] Mark Mero[2]

[1]*School of Analytical Sciences Adlershof, Humboldt-Universität zu Berlin, 12489 Berlin, Germany*
[2]*Max Born Institute for Nonlinear Optics and Short Pulse Spectroscopy, 12489 Berlin, Germany*
*\*Corresponding author: heinerzs@hu-berlin.de*

**We report on a µJ-scale mid-infrared optical parametric amplifier (OPA) based on the recently developed wide-bandgap orthorhombic crystal, BaGa$_4$S$_7$ (BGS), and directly compare its performance to that of LiGaS$_2$ (LGS) in the same OPA setup. The source is based on a single OPA stage amplifying supercontinuum seed pulses with a quantum efficiency of 29% at an idler wavelength of 10 µm, featuring nominally carrier-envelope phase-stable pulses. As a result of pumping the OPA directly at 1 µm, the overall conversion efficiency far exceeds that of traditional schemes based on OPA's followed by difference frequency generation. Chirp compensation using bulk germanium resulted in 126-fs pulses covering the 7.6-11.5-µm spectral range. BGS holds great promise for power scaling due to its availability in larger single-crystal sizes than LGS.**

Generation and amplification of few-cycle laser pulses at repetition rates >> 10 kHz in the short-wave infrared (SWIR, 1.4-3 µm) and mid-infrared (MIR, 3-30 µm) are needed in a growing number of applications ranging from time-resolved vibrational to strong-field spectroscopies and high-harmonic generation [1-6]. A proven route to reaching simultaneously high average and peak power in the SWIR and MIR is through frequency down-conversion of pulses in optical parametric amplifiers (OPA's) from power-scalable solid-state lasers. Compared to laser systems with pulse durations ≥ 10 ps, sub-ps pump sources have the benefit of straightforward supercontinuum-based seed pulse generation for the OPA chain, thereby alleviating the synchronization requirement of the seed and pump pulses. While laser amplifiers in the 2-2.5 µm range are rapidly being developed, the most mature high-average-power ultrafast solid-state laser technology to date is based on diode-pumped (sub-)ps Yb-lasers operating near 1 µm. Such laser amplifiers now routinely deliver kW average powers at peak powers of 10's of GW [7].

Recently, various 1-µm-pumped high-average-power GW-scale MIR OPA's have been demonstrated at center wavelengths below 4 µm [8-10], which relied on commercially available, wide-bandgap oxide nonlinear optical (NLO) crystals, such as LiNbO$_3$, KNbO$_3$, and KTiOAsO$_4$. For several applications, extension of the spectral coverage beyond the IR cutoff of oxide crystals is needed.

There are only a few non-oxide NLO crystals that can be used to generate and amplify pulses above 5 µm, when pumped by ultrafast lasers at 1 µm without detrimental two-photon absorption and associated laser-induced damage [11]. The scarcity of such non-oxide crystals has forced frequency down-conversion schemes to rely on cascaded three-wave mixing with strongly limited conversion efficiencies. In the cascaded scheme, a wide-bandgap oxide crystal-based OPA is followed by a narrow-bandgap non-oxide crystal-based difference frequency generation (DFG) stage. Commercial systems based on cascaded frequency down-conversion of fs Yb laser amplifiers deliver an overall pump-to-MIR energy conversion efficiency of only ≤1.1% in the 4.5-10 µm range [12], where the efficiency peaks at 5 µm and gradually drops at longer wavelengths.

Wide-bandgap non-oxide crystals could potentially allow the direct down-conversion of 1-µm pump pulses to the spectral domain above 5 µm with significantly improved conversion efficiency. However, the development of large, single-domain non-oxide NLO crystals with a UV absorption edge below 500 nm has been hindered by their difficult growth procedures and the necessity of post-growth treatment. The only commercially widely available crystal satisfying the requirements, the chalcopyrite AgGaS$_2$ (AGS), exhibits a bandgap energy of only 2.7 eV resulting in low damage threshold when pumped at 1 µm. Recently, LiGaS$_2$ (LGS) has emerged as a promising material [13,14]. Its room-temperature bandgap energy is 4.15 eV, while its IR absorption edge of ~12 µm is similar to that of AGS [15]. Notwithstanding its relatively low figure of merit compared to AGS, the higher damage resistivity of LGS has enabled the successful demonstration of ultrafast high-repetition-rate MIR OPA's beyond 5 µm at the µJ level [6,16]. Despite this progress, power scaling in LGS is hampered by the limited crystal sizes available using current growth and annealing processes. In contrast, a more recently developed sulfide NLO crystal with the same point group *mm*2 and properties similar to those of LGS, BaGa$_4$S$_7$ (BGS) [17], holds great promise for power scaling, due to the rapid progress in its manufacture allowing single-crystals with larger aperture size [18]. However, BGS has so far not been employed in ultrafast OPA's. Remarkably, when pumped near 1 μm, both LGS and BGS allow broadband phase matching to generate MIR idler pulses in a collinear geometry with almost perfect group velocity matching of all three waves [19,20]. Thus, the generation of broadband MIR idler pulses is possible in long crystals without angular dispersion and significant temporal pump walk-off, leading to efficient depletion of pump radiation. Importantly, the carrier-envelope phase of the generated idler pulses is passively stabilized due to the DFG process between carrier-envelope phase-locked pump and seed pulses. Based on the Sellmeier equations for LGS, the idler wavelength, where the signal and idler group velocities match in type I phase matching in the XZ plane at $\lambda_{pump}$ = 1.028 μm is 7.8 μm [20]. The two sets of Sellmeier equations existing for BGS predict a group velocity-matched idler wavelength at 8.7 μm [19] and 10.4 μm [21].

Here, we present a supercontinuum-seeded, single-stage OPA, where the nonlinear crystal was either LGS or BGS of the same length. Apart from the amplifying medium and the time delay of the chirped supercontinuum seed pulses, all other input pulse parameters were the same to facilitate a direct comparison of the performance of the two crystals. We show that high photon conversion efficiencies are possible using both materials even in the single high-gain amplifier stage configuration. Compared to LGS, BGS showed a higher influence of thermal dephasing possibly due to higher nonlinear refraction and absorption. However, despite the thermal issues, BGS showed higher quantum efficiency than LGS. Similarly to LGS, BGS also supports few-cycle MIR pulse durations with relatively long crystals, interestingly at longer wavelengths than LGS.

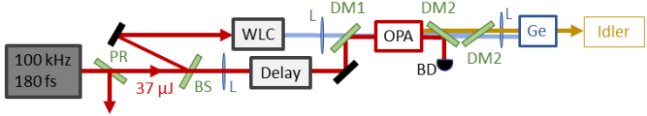

**Fig. 1.** Schematic layout of the tunable MIR OPA. BS: beam sampler, PR: partial reflector, WLC: white light continuum generation unit, L: lens, BD: beam dump. DM1: dichroic mirror, high reflection (HR) at 1.03 μm and high transmission (HT) at >1.1 μm, DM2: dichroic mirror, HR at 1.0-1.2 μm, and HT at 6-12 μm, OPA: LGS or BGS crystal, Ge: germanium-based temporal chirp compensation unit.

Figure 1 shows the scheme of our single-stage MIR OPA. The entire system was driven by a commercial 6 W Yb:KGd(WO$_4$)$_2$ laser system operating at a center wavelength of 1028 nm, a pulse duration of 180 fs, and a repetition rate of 100 kHz. An average power of 3.7 W (i.e., 37 μJ) was used for generating the supercontinuum seed pulses and pumping the OPA. The remaining power was used for other spectroscopic applications [5,6].

For seed pulse generation, a small fraction of the 37-μJ pump pulses was focused into an uncoated, 6-mm-long YAG window. The spectral components of the ultrashort supercontinuum pulses on the Stokes-side of the pump wavelength above 1.06 μm were transmitted through a long-pass filter and refocused using an achromat.

In our OPA, either an 8.0-mm-long LGS or an 8.3-mm-long BGS crystal was employed. Both crystals were uncoated and were cut for type I phase matching in the XZ plane. The performance of the LGS- and BGS-based OPA was compared directly, without changing the focusing conditions. Either the LGS or the BGS sample was placed in the holder of the OPA crystal and the spatial and temporal overlap between the seed and pump pulses were optimized. The supercontinuum seed pulses were focused in front of the OPA crystal through a dichroic mirror (DM1 in Fig. 1) resulting in ~710 μm Gaussian 1/e$^2$ beam diameter at the crystal. The pump pulses were focused using an f = 1000 mm anti-reflection (AR)-coated singlet lens and reflected off the dichroic mirror DM1. The crystals were placed in a diverging pump beam. The pump pulse energy reaching the OPA crystal was up to 34 μJ. The maximum pump peak intensity incident on the front surface of the crystal was up to 45 GW/cm$^2$. The generated MIR idler pulses were separated from the residual pump and the signal pulses using a pair of custom-made dichroic mirrors (cf. DM2 in Fig. 1) and collimated using an AR-coated Ge lens. The maximum MIR average power obtained after the Ge lens was 59 mW at 10 μm and 81 mW at 8.1 μm with the BGS and LGS crystal, respectively. Thus, the overall pump-to-idler energy conversion efficiency, defined as the ratio of the MIR pulse energy after the Ge lens and the total pump pulse energy available at the input of the setup, was 1.6% with BGS and 2.1% with LGS, a factor of ~6 higher than using the OPA + DFG scheme at similar MIR wavelengths.

Figure 2(a) and 2(b) show the pump-to-idler photon conversion efficiency (i.e., quantum efficiency) and energy conversion efficiency of the OPA stage itself, respectively, as a function of pump power. For the data shown in Fig. 2, the losses at both uncoated front and rear crystal interfaces, the two DM2 mirrors, and the Ge lens were corrected for. The measurement was performed at 10 μm and 8.1 μm in the case of BGS and LGS, respectively. While the LGS-based OPA could be pumped at the maximum pump pulse energy of 34 μJ, the BGS-OPA showed signs of strong thermal dephasing above a pump pulse energy of 32 μJ possibly due to multi-photon absorption of the 1-μm pump pulses and/or multi-phonon absorption near 10 μm. Despite the thermal issues in BGS, 29% of photon conversion efficiency was achieved, while its maximum value was 28 % in LGS. The MIR average power reaching the exit face of the crystal was estimated to be 81 mW (i.e., 0.81 μJ) at 10 μm and 105 mW (i.e., 1.05 μJ) at 8.1 for the BGS and LGS OPA, respectively. Both the LGS and BGS single-stage OPA's delivered photon conversion efficiencies on par with multi-stage, supercontinuum-seeded oxide-based MIR OPA's demonstrating the high potential of these wide-bandgap chalcogenide crystals in the 5-10-μm spectral range.

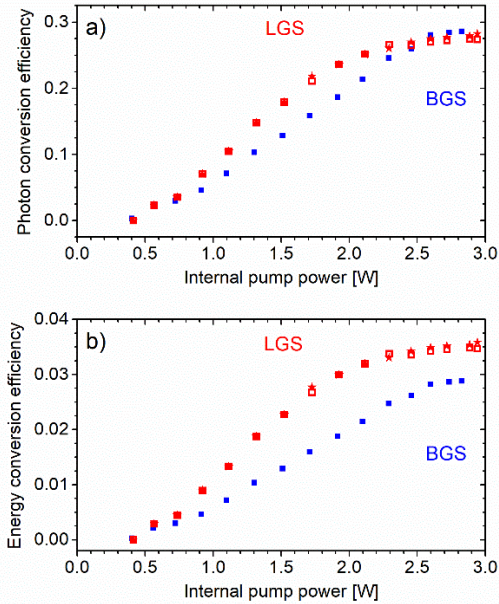

**Fig. 2.** Pump-to-idler photon (a) and energy (b) conversion efficiency as a function of average pump power with a BGS (blue) and an LGS (red) OPA crystal at a center wavelength of 10 and 8.1 µm, respectively. Different symbols of the same color correspond to different independent measurements. The pump power was corrected for Fresnel losses at the entrance interface of the crystal.

The durations of the MIR pulses were measured by a home-built cross-correlation frequency-resolved optical gating (X-FROG) apparatus. For the X-FROG measurements, a 200-µm-long, type I LGS crystal was used in a noncollinear geometry in the XZ principal plane to generate the sum-frequency from the gate and MIR idler pulses. The 180-fs gate pulses at 1.028 µm were delivered by our Yb-pump laser and were characterized separately by a home-built second-harmonic FROG device. The measured and retrieved X-FROG traces together with the reconstructed temporal and spectral profiles of the idler pulses of the BGS-OPA at 10 µm are shown in Figure 3. The idler spectrum was also measured sepa- rately using a commercial Fourier-transform optical spectrum analyzer (cf. green symbols in Fig. 3(d)), showing reasonably good agreement with the reconstructed spectrum from the X-FROG algorithm. The temporal chirp of the MIR idler beam was compensated by transmission through several AR-coated germanium windows with a total length of 49 mm. After temporal chirp compensation, the shortest pulse duration at 10 µm was 126 fs (corresponding to 3.8 optical cycles), which was ~17% longer than the Fourier-limited value (cf. Fig. 3(c)). The shortest obtained pulse duration was limited by uncompensated third-order dispersion.

Similarly short pulses were obtained with the 8-mm-long LGS crystal at 8.1 µm. The corresponding results of the X-FROG measurements are summarized in Fig. 4. After chirp compensation in bulk Ge with a total length of 16 mm, the shortest pulse duration we achieved at 8.1 µm was 121 fs, close to the Fourier-limit (cf. Fig. 4(c)). Here, the idler spectrum was obtained by sum-frequency mixing between the MIR pulses and spectrally narrowband 514-nm pulses at a planar gold surface (cf. green symbols in Fig. 4(d)) [5]. The agreement between the measured and reconstructed spectra was high (cf. Fig. 4(d)).

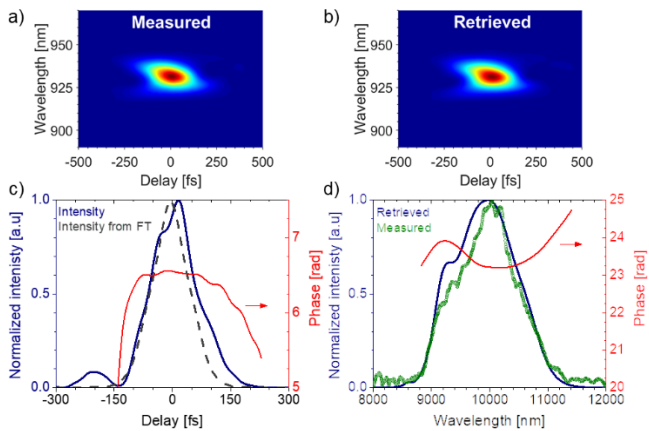

**Fig. 3.** (a) Measured and (b) retrieved X-FROG traces for the chirp-compensated idler pulses generated with an 8.3-mm-long BGS crystal at 10 µm. Reconstructed (c) temporal and (d) spectral intensity and phase. The retrieved pulse duration is 126 fs.

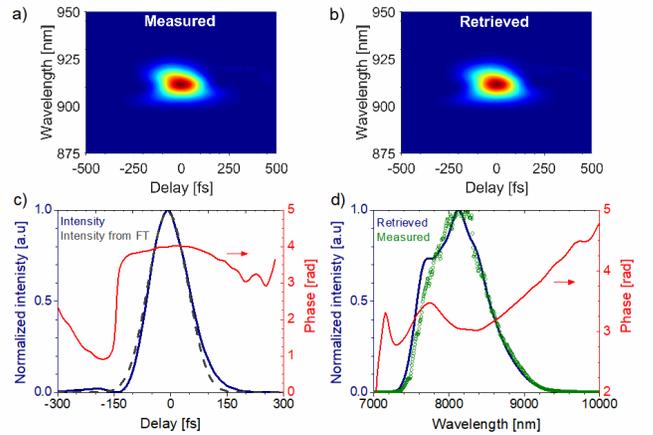

**Fig. 4.** (a) Measured and (b) retrieved X-FROG traces for the chirp-compensated idler pulses generated with an 8-mm-long LGS crystal at 8.1 µm. Reconstructed (c) temporal and (d) spectral intensity and phase. The retrieved pulse duration is 121 fs.

Figure 5(a) shows the spectral tunability of our LGS- and BGS-based OPA's. The wavelength adjustment of the idler pulses was achieved by varying the OPA crystal angle in the XZ plane, followed by the optimization of the spatial and temporal overlap between the pump and seed pulses. The center wavelength of the generated few-cycle, broad-bandwidth MIR idler pulses was tunable between 7.0 and 8.6 µm in the LGS-OPA, and 8.0 µm and 10.9 µm in the BGS-OPA. The wavelengths where the highest pulse energy and the broadest idler bandwidth was reached were approximately the same at 8.1 µm with LGS. In contrast, with BGS, the broadest idler bandwidth was reached at 9.5 µm, while the highest pulse energy was obtained at 10.1 µm. In both cases, the upper limit of the idler wavelength may be due to multi-phonon absorption in the corresponding crystals. For comparison, applying a shorter, 5-mm-long LGS crystal, the center wavelength of the MIR idler pulses could be adjusted in a somewhat wider wavelength range, i.e., between 6 and 9 µm [3], than with the 8.0-mm-long crystal.

Figure 5(b) shows the average power of the MIR idler pulses as a function of center wavelength demonstrating that BGS outperforms LGS in the spectral range above ~8.8 µm. We attribute the relatively narrow tuning ranges to the increasing group velocity mismatch between the three waves away from the optimum wavelength, the IR absorption edges of the crystals.

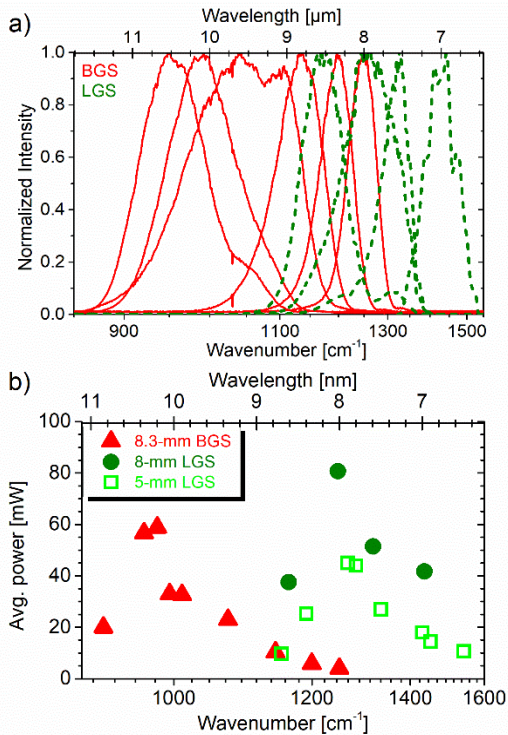

**Fig. 5. (a)** Spectral tunability of the broadband MIR pulses. Solid red lines: BSG-OPA, dashed green lines: LGS-OPA. **(b)** Average power as a function of wavelength in LGS- and BGS-based OPA's. The data values of the 5-mm LGS-OPA are from [6].

In summary, we employed recently developed single-crystal BGS for the generation of few-cycle mid-infrared idler pulses using an OPA pumped by a 1.03-µm Yb-laser. The performance of the BGS crystal in the OPA was directly compared to that of an LGS crystal of the same length. The BGS-OPA operated in the 7.6-11.5-µm spectral range, beyond the 6.5-9.2-µm range of the LGS-OPA. The quantum efficiency in BGS reached 29% at 10 µm, while it peaked at a value of 28% at 8.1 µm in LGS. Starting from a pump power of 3.7 W at a repetition rate of 100 kHz driving the whole setup, the BGS and LGS OPA delivered an output power of 59 mW at 10 µm and 81 mW at 8.1 µm. The overall pump-to-idler energy conversion efficiency of both crystals exceeded that of traditional cascaded frequency down-conversion schemes by a large margin. Higher output powers are feasible using anti-reflection-coated crystals and better optimized coatings on other transmissive optics. Chirp compensation was achieved by transmission through bulk germanium leading to 126-fs pulses with BGS (i.e., 3.8 cycles) and 121-fs pulses with LGS (i.e., 4.5 cycles). The pulse duration with BGS was limited by residual third-order dispersion, which prevented reaching the transform limit of 108 fs. Our results together with the availability of single-crystal BGS in larger sizes than LGS demonstrate that BGS is a promising candidate for power scaling of 1 µm-pumped OPA's beyond the mid-infrared cutoff of oxide crystals.

**Funding.** Horizon 2020 Framework Programme (H2020) (654148); Deutsche Forschungsgemeinschaft (DFG) (GSC 1013 SALSA and PE 607/14-1).

**Acknowledgment**. Z. H. acknowledges funding by a Julia Lermontova Fellowship from DFG (GSC 1013 SALSA).

**References**
1. B. Wolter, M. G. Pullen, M. Baudisch, M. Sclafani, M. Hemmer, A. Senftleben, C. D. Schröter, J. Ullrich, R. Moshammer, and J. Biegert, Phys. Rev. X **5**, 021034 (2015).
2. J. P. Kraack and P. Hamm, Chem. Rev. **117**, 10623 (2017).
3. X. Ren, J. Li, Y. Yin, K. Zhao, A. Chew, Y. Wang, S. Hu, Y. Cheng, E. Cunningham, Y. Wu, M. Chini, and Z. Chang, J. Opt. **20**, 023001 (2018).
4. A. Ge, B. Rudshteyn, P. E. Videla, C. J. Miller, C. P. Kubiak, V. S. Batista, and T. Lian, Acc. Chem. Res. **52**, 1289 (2019).
5. Z. Heiner, V. Petrov, and M. Mero, APL Photonics **2**, 066102 (2017).
6. Z. Heiner, L. Wang, V. Petrov, and M. Mero, Opt. Express **27**, 15289 (2019).
7. P. Russbueldt, D. Hoffmann, M. Höfer, J. Löhring, J. Luttmann, A. Meissner, J. Weitenberg, M. Traub, T. Sartorius, D. Esser, R. Wester, P. Loosen, and R. Poprawe, IEEE J. Sel. Top. Quantum Electron. **21**, 447 (2015).
8. U. Elu, M. Baudisch, H. Pires, F. Tani, M. H. Frosz, F. Köttig, A. Ermolov, P. St.J. Russell, and J. Biegert, Optica **4**, 1024 (2017).
9. M. Mero, Z. Heiner, V. Petrov, H. Rottke, F. Branchi, G. M. Thomas, and M. J. J. Vrakking, Opt. Lett. **43**, 5246 (2018).
10. N. Thiré, R. Maksimenka, B. Kiss, C. Ferchaud, G. Gitzinger, T. Pinoteau, H. Jousselin, S. Jarosch, P. Bizouard, V. Di Pietro, E. Cormier, K. Osvay, and N. Forget, Opt. Express **26**, 26907 (2018).
11. V. Petrov, Prog. Quantum Electron. **42**, 1 (2015).
12. P. M. Donaldson, G. M. Greetham, D. J. Shaw, A. W. Parker, and M. Towrie, J. Phys. Chem. A **122**, 780 (2018).
13. L. Isaenko, A. Yelisseyev, S. Lobanov, P. Krinitsin, V. Petrov, and J. J. Zondy, J. Non-Crystal. Solids **352**, 2439 (2006).
14. L. I. Isaenko, and I. G. Vasilyeva, J. Cryst. Growth **310**, 1954 (2008).
15. A. Yelisseyev, Z. S. Lin, M. Starikova, L. Isaenko, and S. Lobanov, J. Appl. Phys. **111**, 113507 (2012).
16. S. Qu, H. Liang, K. Liu, X. Zou, W. Li, Q. J. Wang, and Y. Zhang, Opt. Lett. **44**, 2422 (2019).
17. X. Lin, G. Zhang, and N. Ye, Cryst. Growth Des. **9**, 1186 (2009).
18. V. Badikov, D. Badikov, G. Shevyrdyaeva, A. Tyazhev, G. Marchev, V. Panyutin, F. Noack, V. Petrov, and A. Kwasniewski", Opt. Mater. Express **1**, 316 (2011).
19. N. Umemura, V. Petrov, V. Badikov, and K. Kato, Proc. SPIE **9347**. 93470N (2015).
20. K. Kato, K. Miyata, L. Isaenko, S. Lobanov, V. Vedenyapin, and V. Petrov, Opt. Lett. **42**, 4363 (2017).
21. V. Badikov, D. Badikov, G. Shevyrdyaeva, A. Tyazhev, G. Marchev, V. Panyutin, V. Petrov, and A. Kwasniewski, Phys. Stat. Sol. RRL **5**, 31 (2011).